\documentstyle[psfig,conf_iap,10pt]{article}

\begin{document}

\heading{%
The Universe on Very Large Scales:  \\
A View from the Las Campanas Redshift Survey} 
\par\medskip\noindent

\author{%
Douglas L.\ Tucker$^{1}$, Huan Lin$^{2}$, Stephen Shectman$^{3}$
}
\address{%
Fermilab, MS~127, P.O. Box 500, Batavia, IL 60510, USA.
}
\address{%
Steward Observatory, University of Arizona, Tucson, AZ 85721, USA.
}
\address{%
Carnegie Observatories, 813 Santa Barbara St., Pasadena, CA 91101, USA.
}

\begin{abstract}
The Las Campanas Redshift Survey (LCRS) is among the first galaxy
redshift surveys to sample a reasonably fair volume of the local
Universe.  On the largest scales ($\gg 100~h^{-1}$~Mpc), the galaxy
distribution appears smooth; on relatively small scales ($\lsim
10~h^{-1}$~Mpc), the LCRS tends to confirm the clustering
characteristics observed in previous, shallower surveys.  Here,
however, we concern ourselves primarily with clustering on scales near
the transition to homogeneity (50 -- 200$~h^{-1}$~Mpc).  We conclude
that the general evidence tends to support enhanced clustering on
$\sim 100~h^{-1}$~Mpc scales, but that this result should be confirmed
with additional analyses of the LCRS dataset (especially 2D analyses)
and with investigations of new and upcoming large-scale surveys
covering different regions and/or having different selection effects.
\end{abstract}

\section{Introduction}

When the strategy for the Las Campanas Redshift Survey (LCRS;
\cite{Shectman96}) was first being formulated in the mid-1980's, the
goals were two-fold: first, to sample a fair volume of the local
Universe (what are the largest coherent structures?  at what point
does the Universe start to look smooth?), and, second, to study
clustering on large scales ($\xi_{\rm gg}(s)$, void statistics, ...).
Therefore, it was decided to survey an unfilled ``checkerboard''
pattern in both the north and the south galactic caps out to a
redshift of $z \sim 0.2$.  To facilitate these ends, it was decided to
use a 50-fiber multi-object spectrograph (MOS) to obtain the radial
velocities.  This was circa 1986/87.  A few years into the survey
(circa 1990/91), the survey strategy evolved into what became its
final form: a set of six filled-in slices -- three in the north
galactic cap, and three in the south.  Also during this time frame,
the MOS was upgraded to 112-fibers.

The maps that were gradually built up over the course of the survey
eventually began to show an intriguing picture -- that of a Universe
which looked smooth on large scales!  Visually, there was little or no
evidence in the LCRS slices for high-contrast coherent structures on
scales $\gg 100~h^{-1}$~Mpc.  Quantitative evidence has tended to
support this initial view (e.g., see Fig.~8 of \cite{Lin96a} and
Fig.~1 of \cite{Tucker97}).

At the other extreme, on relatively small scales ($\lsim
10~h^{-1}$~Mpc), the LCRS has generally confirmed what had been
observed in previous, shallower surveys (\cite{Lin96b},
\cite{Tucker97}).

But what about scales near the transition to homogeneity (50 --
200$~h^{-1}$~Mpc)?  What can the LCRS tell us about clustering on
these scales?

\section{Results}

\begin{figure}[t]
\centerline{\vbox{
\psfig{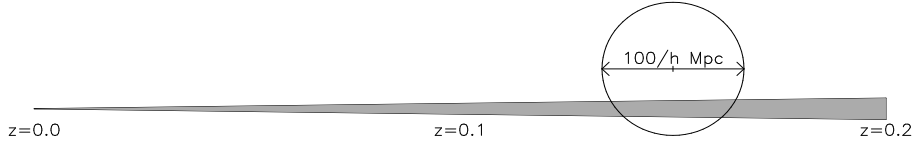}
}}
\caption[]{A schematic of the edge-on view of an LCRS slice (drawn to
scale); at $z=0.2$, the width of an LCRS slice has grown to $\approx
15~h^{-1}$~Mpc.  Also drawn to scale is a $100~h^{-1}$~Mpc-diameter
sphere randomly intersecting the LCRS slice.}
\end{figure}

On very large scales ($\gsim 50~h^{-1}$~Mpc), it makes sense to think
of an individual LCRS slice as a 2-dimensional plane, since the third
dimension of slice width does not contribute much to our understanding
of clustering on these scales (Fig.~1).  Now consider a ``toy''
Universe composed of osculating spherical voids of diameter
$100~h^{-1}$~Mpc.  In such a Universe, an LCRS slice would intersect
several voids at various random positions.  Clearly, the measured
cross-sectional diameters of these intersections will never
overestimate the true diameter of the spheres (in this case,
$100~h^{-1}$~Mpc).  On average, the diameter of the cross-section of a
sphere randomly intersecting a plane is
\begin{equation}
<D_{\rm cross-section}> =       \frac{\pi}{4} D_{\rm sphere} 
			\approx \frac{3}{4}   D_{\rm sphere} .
\end{equation}
Thus, in such a ``toy'' Universe, we would expect that an LCRS slice
would typically underestimate the average void diameter by about
$1/3$.  This {\em projection effect\/} in real space is basically the
same as the {\em aliasing\/} of power from large- to small-scales in
Fourier space.  In real-space, however, this effect seems rather
more benign and correctable.

With the above discussion in mind, what -- if any -- clustering do we
see on scales of 50 -- 200$~h^{-1}$~Mpc?

\begin{figure}
\centerline{\vbox{
\psfig{figure=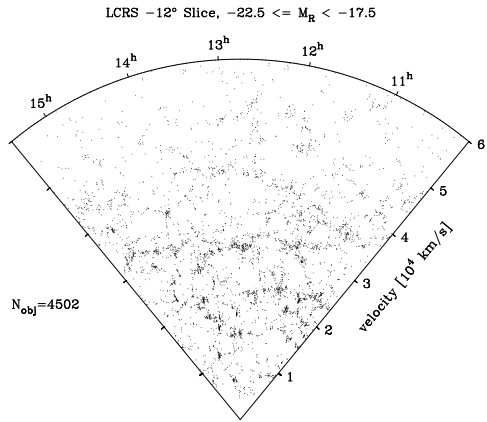,height=7.0cm,angle=-90.}
}}
\caption[]{The LCRS Slice at $-12^{\circ}$.  Only LCRS galaxies in
the official redshift sample and within the absolute magnitude range
$-22.5 \le M_{\rm R} - 5\log h < -17.5$ are plotted.}
\end{figure}
\begin{figure}
\centerline{\vbox{
\psfig{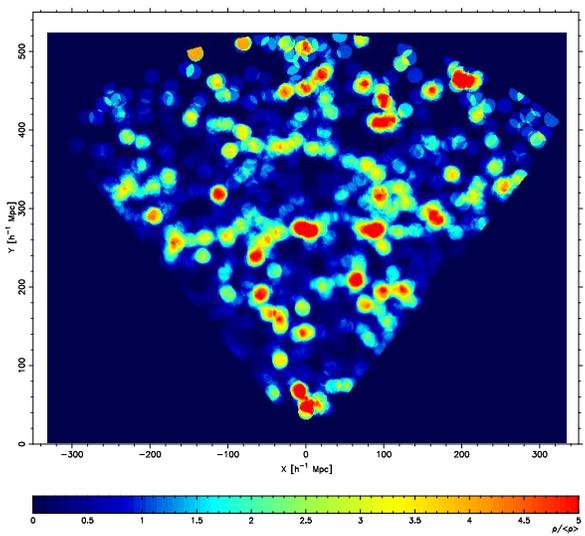}
}}
\caption[]{Smoothed density contrast map of the LCRS $-12^{\circ}$
Slice.  Smoothing was performed with a $20~h^{-1}$~Mpc-diameter tophat
filter.  The $xy$-coordinates are in comoving distances.}
\end{figure}

First, consider a couple visual representations of the LCRS
$-12^{\circ}$ Slice, the most densely and most homogeneously sampled
of the six LCRS slices.  Looking at a standard velocity-RA plot of
this slice (Fig.~2), we may notice that the largest coherent
high-contrast structures tend to form the walls of underdense regions
(``voids'') 50 -- 100$~h^{-1}$~Mpc in diameter.  To remove radial and
field-to-field selection effects, one can generate a smoothed
number-density contrast map (Fig.~3); therein, our initial suspicions
are confirmed: high-density ($\delta\rho/\rho \gsim 2.5$) regions tend
to surround low-density ($\delta\rho/\rho \lsim 1$) ``voids'' with
diameters of 50 -- 100$~h^{-1}$~Mpc.  

Second, we note that both the LCRS 2D power spectrum \cite{Landy96}
and large-scale ($>10~h^{-1}$~Mpc) 3D spatial autocorrelation function
\cite{Tucker97} indicate features of excess clustering on scales $\sim
100~h^{-1}$~Mpc.  We note that the results from the 2D power spectrum
show higher statistical significance, since the signal at these scales
(see Fig.~1) is essentially 2D; at these scales, some of the signal is
washed out in the 3D autocorrelation function analysis.

Third, Doroshkevich et al.'s core-sampling analysis
\cite{Doroshkevich96} of the LCRS measures the mean free path between
2D sheets and between 1D filaments.  Comparing \cite{Doroshkevich96}'s
Fig.~12 with our Fig.~3, it is apparent that their sheet-like
structures correspond roughly with regions of smoothed
$\delta\rho/\rho \gsim 2.5$; their rich filaments, with regions of
smoothed $1.5 \lsim \delta\rho/\rho \lsim 2$.  Doroshkevich et al.\ 
measure the mean free path between sheets to be $\approx$ 80 --
100~$h^{-1}$~Mpc \cite{Doroshkevich96}.

Fourth, Einasto et al.\ give evidence that Abell Clusters in rich
superclusters tend to lie within a 3D ``chessboard'' of gridsize of
$\sim 120~h^{-1}$~Mpc \cite{Einasto97}.  Einasto and his colleagues
are presently performing a similar analysis for LCRS galaxies in rich
environments \cite{Einasto98}; in fact, our Fig.~3 is from the early
stages of that analysis.  Results, however, are still pending.

\section{Conclusions}

There does seem to be something going on in the LCRS at a scale of
$\sim 100~h^{-1}$~Mpc, but this result needs confirmation using other
techniques (e.g., a 2D $\xi_{\rm gg}$) and using other large surveys
covering different regions and/or having different selection effects
(e.g., the ESP \cite{ESP}, 2dF \cite{2dF}, SDSS \cite{SDSS}, ...).

\acknowledgements{We wish to thank the following for fruitful
discussions regarding the topic of this paper: Andrei Doroshkevich,
Jaan Einasto, Richard Fong, Yasuhiro Hashimoto, Robert Kirshner,
Stephen Landy, Augustus Oemler, and Paul Schechter.}

%References should be refered as : \cite{LH}, \cite{MMM}, and \cite{Kea}. 

\vfill
\end{document}